\documentclass[conference]{IEEEtran}
\IEEEoverridecommandlockouts
\usepackage{cite}
\usepackage{amsmath,amssymb,amsfonts}
\usepackage{graphicx}
\usepackage{textcomp}
\usepackage{xcolor}
\def\BibTeX{{\rm B\kern-.05em{\sc i\kern-.025em b}\kern-.08em
    T\kern-.1667em\lower.7ex\hbox{E}\kern-.125emX}}

\usepackage{hyperref}
\usepackage[nolist,nohyperlinks]{acronym}
\usepackage{wrapfig}
\usepackage{todonotes}
\usepackage{tikz}
\usetikzlibrary{fit,shapes,arrows,automata}
\usepackage[caption=false]{subfig}

\usepackage{centernot}
\usepackage{cancel}
\usepackage{amsmath}
\usepackage{amssymb}
\usepackage{amsthm}
\usepackage{mathtools}
\usepackage{makecell}
\usepackage{algorithm}
\usepackage[noend]{algpseudocode}
\usepackage{enumitem}
\usepackage{trimclip}

\theoremstyle{definition}
\newtheorem{example}{Example}[section]

\IEEEtriggeratref{28}
\IEEEtriggercmd{\enlargethispage{-4.5cm}}
\begin{document}
\begin{acronym}
\acro{TTS}{timed transition system}
\acro{SUT}{system under test}
\acro{CAS}{car alarm system}
\acro{PC}{particle counter}

\acroplural{SUT}[SUTs]{systems under test}

\acro{TA}{timed automaton}
\newacroplural{TA}[TA]{timed automata}

\end{acronym}
\newcommand{\tap}{\acp{TA}}
\newcommand{\ta}{\ac{TA}}

\newcommand{\twodots}{\mathinner {\ldotp \ldotp}}

\newcommand{\exec}{\textsc{exec}}
\newcommand{\verdict}{\textsc{verdict}}
\newcommand{\out}{\textsc{out}}
\newcommand{\step}{\textsc{step}}

\newcommand{\fail}{\mathtt{FAIL}}
\newcommand{\nondet}{\mathtt{NONDET}}
\newcommand{\pass}{\mathtt{PASS}}

\newcommand{\ntest}{n_\mathrm{test}}
\newcommand{\ptest}{p_\mathrm{test}}
\newcommand{\pcross}{p_\mathrm{cr}}

\newcommand{\npop}{n_\mathrm{pop}}
\newcommand{\nmig}{n_\mathrm{mig}}
\newcommand{\nsel}{n_\mathrm{sel}}
\newcommand{\gmax}{g_\mathrm{max}}
\newcommand{\gchange}{g_\mathrm{change}}
\newcommand{\gsimp}{g_\mathrm{simp}}

\newcommand{\nloc}{n_\mathrm{loc}}
\newcommand{\nclock}{n_\mathrm{clock}}
\newcommand{\cmax}{c_\mathrm{max}}
\newcommand{\pmut}{p_\mathrm{mut}}
\newcommand{\pmutinit}{{p_\mathrm{mut}}_\mathrm{init}}

\newcommand{\ntourn}{n_\mathrm{t}}
\newcommand{\ptourn}{p_\mathrm{t}}

\title{Learning Timed Automata\\ via Genetic Programming
}

\author{\IEEEauthorblockN{Martin Tappler,\hspace{5mm}
    Bernhard K. Aichernig}
\IEEEauthorblockA{\textit{Institute of Software Technology} \\
\textit{Graz University of Technology, Austria}\\
%author1@anon.org
}
\and
\IEEEauthorblockN{Kim Guldstrand Larsen, \hspace{5mm} Florian Lorber}
\IEEEauthorblockA{\textit{Department of Computer Science} \\
\textit{Aalborg University, Denmark}}
}

\maketitle

\begin{abstract}
Model learning has gained increasing interest in recent years. It derives behavioural models from test data of black-box systems. The main advantage offered by such techniques is that they enable model-based analysis without access to the internals of a system. Applications range from fully automated testing over model checking to system understanding. Current work focuses on learning variations of finite state machines. However, most techniques consider discrete time. In this paper, we present a method for learning timed automata, finite state machines extended with real-valued clocks. The learning method generates a model consistent with a set of timed traces collected by testing. This generation is based on genetic programming, a search-based technique for automatic program creation. We evaluate our approach on $\mathbf{44}$ timed systems, comprising four systems from the literature and $\mathbf{40}$ randomly generated examples.
\end{abstract}


\section{Introduction}
Test-based model-learning techniques have gained increasing interest in recent years.
Basically, these techniques derive formal system models from (test) observations.
They therefore enable model-based reasoning about software systems 
while requiring only limited knowledge about the system at hand. 
Put differently, such techniques allow for model-based verification of black-box
systems if they are amenable to testing.

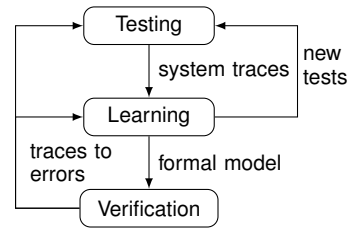
\begin{figure}[t]
\centering%
 \begin{tikzpicture}[font=\footnotesize\sffamily,text width = 1.5cm,text centered, minimum height = 0.5cm, node distance = 0.7cm]%
  \node[rounded corners, draw](test){Testing};
  \node[rounded corners, draw, below = of test](learn){Learning};
  \node[rounded corners, draw, below = of learn,text width = 1.6cm](verif){Verification};
%   \node[below = of eval](result){};
  \draw[-latex] (test) --node[right=0cm, align=left, text width = 5em]{system traces} (learn);
  \draw[-latex] (learn) --node[right=0cm, align=left, text width = 5em](form_model){formal model} (verif);
  
%   \node[rounded corners, draw, left = 1cm of learn_to_reach](sample_sched){Sample with Strategy};
  
  \draw[-latex] (learn.east) --+(1.1cm,0)|-node[above right = -0.9cm and -0.05cm, align=left,text width = width("tests")]{new tests} (test.east);
  \draw[-latex] (verif.west) --+(-0.85cm,0)|- (learn.west);
  \draw[-latex] (verif.west) --+(-0.85cm,0)|- (test.west);
  \node[left = -0.3cm of form_model, align=left,text width = 5em]{traces to errors};
  
 \end{tikzpicture}%
 \caption{General framework for test-based learning enabling verification.}%
 \label{fig:intro_overview}
\end{figure}

Peled et al.~\cite{Peled2002} performed pioneering work in this area by introducing
\emph{Black Box Checking}, automata-based model checking for black-box systems. It 
involves interleaved model learning, model checking and conformance testing
and built the basis for various follow-up work~\cite{Groce_et_al_2002,Elkind_et_al_2006}.
More recent work in this area includes for example
model checking of network protocols~\cite{DBLP:conf/cav/Fiterau-Brostean16,DBLP:conf/spin/Fiterau-Brostean17}, and
differential testing on the model level~\cite{DBLP:conf/sp/SivakornAPKJ17,DBLP:conf/ccs/ArgyrosSJKK16,DBLP:conf/icst/TapplerAB17}.
The basic framework we target is shown in Fig.~\ref{fig:intro_overview}. In the simplest case, we 
interact with a system by testing, learn a model from system traces and then perform some kind of verification.
However, feedback loops are possible: we can derive additional tests from the preliminary learned model, and we could use 
counterexample traces from model checking as tests.%may also be used as tests.

% \setlength{\intextsep}{0pt}%
% \begin{wrapfigure}[9]{t}{4.35cm}%

Learning-based verification has great potential, but applications often use 
modelling formalisms with low expressiveness such as Mealy machines. 
This can be attributed to the availability of efficient implementations of learning algorithms for variations of finite automata, 
% e.g. in LearnLib~\cite{isberner2015}, and comparably low support for richer types of automata.
% In particular, timed systems have received little attention. Notable works include 
e.g. in LearnLib~\cite{isberner2015}, and comparably low support for richer automata types;
especially timed systems have received little attention. Notable works include 
learning of real-time automata and a probabilistic variant thereof~\cite{verwer2007algorithm,DBLP:conf/icgi/VerwerWW10} by Verwer et al.
and  techniques for learning event-recording automata described by Grinchtein et al.~\cite{DBLP:journals/tcs/GrinchteinJL10,DBLP:conf/concur/GrinchteinJP06}.
Recently, Jonsson and Vaandrager also developed an active learning technique for Mealy machines
with timers~\cite{mealy_timer}. 
Our goal is to overcome limitations of these approaches. Real-time automata, e.g., are restricted in their expressiveness
and learning of event-recording automata has a high runtime complexity.  

\paragraph*{Scope and Outline}

In this work, we focus on the \emph{learning} part in Fig.~\ref{fig:intro_overview}. 
Generally, model learning may be performed either passively or actively~\cite{delaHiguera_2010}.
In the former, preexisting data, such as system logs or existing test data, serves as a basis, while 
in the latter, the system is actively queried during the testing phase, i.e. tested, to gain relevant information. 
The technique we propose is passive in general, but active extensions are possible.

More specifically, we use a form of genetic programming~\cite{DBLP:books/daglib/0070933} to
automatically learn a deterministic \ta{} consistent with a given set of test cases. 
Our main contribution is the development of a genetic-programming framework for \tap{}
which includes mutation operators, % targeting timed systems, 
a crossover procedure,
and a corresponding, fine-grained fitness-evaluation. 
We evaluate this approach, a meta-heuristic search, on four manually created \tap{} and 
several randomly generated \tap{}. The evaluation demonstrates that the 
search reliably converges to a \ta{} consistent with the 
test cases given as training data. Furthermore, we simulate each generated \ta{} on independently 
produced test data to show that our identified solutions generalise well, thus
do not overfit to training data. 

% : a car alarm system~\cite{DBLP:conf/tap/AichernigLN13},
% and a \emph{train gate controller}~\cite{DBLP:conf/forte/YiPD94}. 
% The former demonstrates that human-readable models can be learned, while
% the latter demonstrates usage in a verification context.

This paper is structured as follows. The next section contains background information on \tap{}
and genetic programming. Section~\ref{sec:gen_prog} describes our approach to learning \tap{}.
Applications of this approach are presented in Sect.~\ref{sec:case_studies}. In Sect.~\ref{sec:conc}, 
we provide a summary and discuss related work, as well as potential extensions.
% discuss related work in the fields of model learning and genetic programming and
% discuss potential extensions. 

\section{Preliminaries}
\subsection{Timed Automata}
\label{sec:ta_prelim}
Timed automata are finite automata enriched with real-valued variables called clocks~\cite{DBLP:journals/tcs/AlurD94}.
Clocks measure the progress of time which elapses while an automaton resides in some location.
Transitions can be constrained based on clock values and clocks may be reset upon the execution of transitions.
We denote the set of clocks by $\mathcal{C}$ and the set of guards over $\mathcal{C}$ by $\mathcal{G}(\mathcal{C})$.
Guards are conjunctions of constraints of the form $c \oplus k$, with $c \in \mathcal{C}, \oplus \in \{>, \geq, \leq, <\}, k \in \mathbb{N}$.
Transitions are labelled by input and output actions, denoted by $\Sigma_I$
and $\Sigma_O$ respectively, with $\Sigma = \Sigma_I \cup \Sigma_O$. Input action labels are suffixed by $?$, while
output labels contain the suffix $!$. A timed automaton over $(\mathcal{C},\Sigma)$ is a triple $\langle L, l_0, E \rangle$, where $L$
is a finite non-empty set of locations, $l_0 \in L$ is the initial location and $E$ is the set of edges, with 
$E \subseteq L \times \Sigma \times \mathcal{G}(\mathcal{C}) \times 2^\mathcal{C} \times L$. We write 
$l \xrightarrow{g,a,r} l'$ for an edge $(l,g,a,r,l') \in E$ with a guard $g$, an action label $a$, and clock resets $r$. 
\begin{example}[Train \tap{} Model]
Figure~\ref{fig:learned_train} shows a \ta{} model of a train, 
for which we have $\Sigma_I = \{\mathit{start}?,\mathit{stop}?,\mathit{go}?\}$, 
$\Sigma_O = \{\mathit{appr}!,\mathit{enter}!,\mathit{leave}!\}$, $\mathcal{C} = \{c\}$, $L = \{l_0,\ldots,l_5\}$,
and $E = \{l_0 \xrightarrow{\top,\mathit{start}?,\{c\}} l_1, \ldots\}$.
From initial location $l_0$, the train accepts the input $\mathit{start}?$, resetting clock $c$. % denoted by $c$ in curly braces.
After that, it can produce the output $\mathit{appr}!$ if $c \geq 5$, i.e the train may approach  
$5$ time units after it is started. %We will discuss this in more detail in Sect.~\ref{sec:case_studies}.
%s5 \xrightarrow{c \geq 5,appr!,\{c\}} s3, s3 \xrightarrow{true,stop?,\{c\}} s2, s3 \xrightarrow{c \geq 10,enter!,\{c\}} s4, s2 \xrightarrow{true,go?,\{c\}} s7,$\\$ s7 
%\xrightarrow{c \geq 7,enter!,\{c\}} s4 , s4 \xrightarrow{c \geq 3,leave!,\{\}} s0\}$. 
\end{example}
% TODO: check again whether we still need that
% We denote sequences of actions $a_1,\ldots, a_n$ by $a_1 \cdots a_n$. The concatenation of sequences $s$ and $u$ is $s \cdot u$
% and a single action $a$ is implicitly lifted to a sequence of length one. The empty sequence is $\epsilon$ and if 
% $S$ and $U$ are sets of sequences, then $S \cdot U = \{s \cdot u| s \in S, u \in U\}$ are pairwise concatenations 
% of their elements. We use $u \ll s = \exists u' : u\cdot u' = s$ to denote that $u$ is a prefix of $s$.

% \begin{wrapfigure}[10]{tb}{4.2cm}%
\begin{figure}[t]%
\centering%
\begin{tikzpicture}[>=stealth',shorten >=1pt,auto,node distance=1.5cm, every node/.append style={font=\scriptsize},
    every edge/.append style={nodes={font=\scriptsize}},every state/.style={minimum size=0pt,inner sep =1pt}]
   \node[state] (l_0)      {$l_0$};
   \node[right = 0.3cm of l_0] (init)      {};
  
  \node[state] (l_1) [left = of l_0] {$l_1$};
  \node[state] (l_2) [below = 0.6cm of l_1] {$l_2$};
  \node[state] (l_3) [below = 0.6cm of l_2] {$l_3$};
  \node[state] (l_4) [below = 0.6cm of l_3] {$l_4$};
  \node[state] at(l_4 -| l_0) (l_5) {$l_5$};
  
  \draw[->](l_0) --node[below,text width = 0.8cm]{$\top$ $\mathit{start}?$ $\{c\}$ } (l_1);
  \draw[->](l_1) --node[left=0.2cm,text width = 0.8cm,align=right]{$c\geq5$ $\mathit{appr}!$ $\{c\}$ } (l_2);
  \draw[->](l_2) --node[left=0.2cm,text width = 0.8cm,align=right]{$\top$ $\mathit{stop}?$ $\{c\}$ } (l_3);
  \draw[->](l_2) --node[above right=0.1cm and -0.2cm,text width = 0.8cm]{$c\geq10$ $\mathit{enter}!$ $\{c\}$ } (l_5);

  \draw[->](l_3) --node[left=0.2cm,text width = 0.6cm,,align=right]{$\top$ $\mathit{go}?$ $\{c\}$ } (l_4);
  \draw[->](l_4) --node[above left=0cm and -0.3cm,text width = 0.8cm]{$c\geq7$ $\mathit{enter}!$ $\{c\}$ } (l_5);
  \draw[->](l_5) --node[right,text width = 0.8cm]{$c\geq 3$ $\mathit{leave}!$ $\{\}$ } (l_0);
  
  \draw[->] (init) -- (l_0);
\end{tikzpicture}
% \vspace{-0.7cm}
 \caption{Train \ta{}.}
\label{fig:learned_train}
\end{figure}
%The semantics of a \ta{} is given in terms of a \ac{TTS} $\langle Q,q_0, \Sigma, T \rangle$, with 
The semantics of a \ta{} is given by a \ac{TTS} $\langle Q,q_0, \Sigma, T \rangle$, with 
states $Q = L \times {\mathbb{R}_{\geq 0}}^\mathcal{C}$, initial state $q_0$, and transitions $T \subseteq Q \times (\Sigma \cup \mathbb{R}_{\geq 0}) \times Q$,
for which we write $q~\xrightarrow{e}~q'$ for $(q,e,q') \in T$.
A state $q = (l, \nu)$ is a pair consisting of a location $l$ and a clock valuation $\nu$. For $r \subseteq \mathcal{C}$, we
denote resets of clocks in $r$ by $\nu[r]$, i.e. $\forall c \in r : \nu[r](c) = 0$ and $\forall c \in \mathcal{C} \setminus r : \nu[r](c) = \nu(c)$.
Let $(\nu + d)(c) = \nu(c) + d$ for $d \in \mathbb{R}_{\geq 0}, c \in \mathcal{C}$ denote the progress of time and 
$\nu \models \phi$ denote that valuation $\nu$ satisfies formula $\phi$. Finally, $\mathbf{0}$ is the valuation assigning zero
to all clocks and the initial state $q_0$ is $(l_0, \mathbf{0})$. 
Transitions of \acp{TTS} are either delay transitions $(l,\nu) \xrightarrow{d} (l,\nu + d)$ for a delay $d \in \mathbb{R}_{\geq0}$, 
or discrete transitions $(l,\nu) \xrightarrow{a} (l', \nu[r])$ for an edge $l \xrightarrow{g,a,r} l'$ such that $\nu \models g$.
Delays are usually further constrained, e.g. by invariants~\cite{DBLP:conf/fates/HesselLNPS03} limiting the sojourn time 
in locations. %We will discuss this below. 

\emph{Timed Traces.}
We use the terms timed traces and test sequences similarly to~\cite{DBLP:journals/tcs/SpringintveldVD01}.
The latter are sequences of inputs and corresponding execution times, 
while the former are sequences of inputs and outputs, together with their times of occurrence (produced in response to a test sequence).
A test sequence $\mathit{ts}$ is an alternating sequence of non-decreasing time stamps $t_j$ and inputs $i_j$, i.e.
$\mathit{ts} = t_1 \cdot i_1 \cdots t_n \cdot i_n \in (\mathbb{R}_{\geq0} \times \Sigma_I)^*$ with $\forall j \in \{1,\ldots, n-1\} : t_j \leq t_{j+1}$.
Informally, a test sequence prescribes that $i_j$ should be executed at time $t_j$.
A timed trace $\mathit{tt} \in (\mathbb{R}_{\geq0} \times \Sigma)^*$ consists of inputs interleaved with outputs produced by a timed system.
% a timed automaton or the \ac{SUT} which can be modelled by a timed automaton.
Analogously to test sequences, its timestamps are non-decreasing. 

\subsubsection{Assumptions on Timed Systems}

Testing based on \tap{} often places further assumptions on \tap{}~\cite{DBLP:journals/tcs/SpringintveldVD01,DBLP:conf/fates/HesselLNPS03}.
Since we learn models from tests we make similar assumptions (closely following \cite{DBLP:conf/fates/HesselLNPS03}). We describe these assumptions on the level of semantics 
and use $q \xrightarrow{a}$ to denote $\exists q' : q \xrightarrow{a} q'$ and $q \centernot{\xrightarrow{a}}$ for $\nexists q' : q \xrightarrow{a} q'$:
%and $q \xRightarrow{a_1 \cdots a_n} q'$ if $\exists q_1, \ldots, q_{n+1}: q_i \xrightarrow{a_i} q_{i+1}$ with $q_1 = q$ and $q_{n+1} =q'$.
% \todo{we do this to ensure controllability and to approximate equivalence checking via testing}
\begin{enumerate}
 \item {\sl Determinism.}
 A \ta{} is deterministic iff for every state $s = (l, \nu)$ and every action $a \in \Sigma$, whenever 
 $s \xrightarrow{a} s'$, and $s \xrightarrow{a} s''$ then $s' = s''$.
 \item {\sl Input Enabledness.}
 A \ta{} is input enabled iff for every state $s = (l, \nu)$ and every input $i \in \Sigma_I$, 
 we have $s \xrightarrow{i}$. % whenever $s \xrightarrow{d}$ for $d \in \mathbb{R}_{\geq0}$.
 \item {\sl Output Urgency.}
 A \ta{} shows output-urgent behaviour if outputs occur immediately as soon as they are enabled, i.e.
 for $o \in \Sigma_O$, if $s \xrightarrow{o}$ then $s \centernot{\xrightarrow{d}}$ for all $d \in \mathbb{R}_{\geq0}$.
 I.e., outputs must not be delayed.
 \item{\sl Isolated Outputs.}
 A \ta{} has isolated outputs iff whenever an output may be executed, then no other output is enabled, i.e.
 if $\forall o \in \Sigma_O, \forall o' \in\Sigma_O$ $q \xrightarrow{o}$ and $q\xrightarrow{o'}$ implies $o = o'$.
\end{enumerate}

It is necessary to place restrictions on the sojourn time in locations to establish output urgency. 
Deadlines provide a simple way to model the assumption that systems are output urgent~\cite{DBLP:conf/compos/BornotST97}.
With deadlines it is possible to model eager actions.
We use this concept and implicitly assume all learned output edges to be eager. This means
that outputs must be produced as soon as their guards are satisfied. For that, we extend the semantics 
% given above by adding the following restriction: delay transitions $(l,\nu) \xrightarrow{d} (l',\nu + d)$ are only possible if 
% $\forall d' \in \mathbb{R}_{\geq 0}, d' < d$ we have $\nu + d' \models \bigvee_{g \in G_O} \lnot g$, 
given above by adding the following restriction: delays $(l,\nu) \xrightarrow{d} (l',\nu + d)$ are only possible if 
$\forall~d'~\in~\mathbb{R}_{\geq 0}, d' < d : \nu + d' \models \lnot \bigvee_{g \in G_O} g$, 
where $G_O = \{g | \exists l', a, r: l\xrightarrow{g,a,r}~l', a\in\Sigma_O \}$ are the guards of outputs in location $l$.
To avoid issues related to the exact time at which outputs should be produced, we further restrict the syntax of \tap{} by disallowing 
strict lower bounds for output edges.
% (consider for instance reaching a location with a valuation $\nu = \{c \mapsto 0\}$ 
% which defines an output with guard $c > 0$). 
\textsc{Uppaal}~\cite{DBLP:journals/sttt/LarsenPY97} uses invariants rather than deadlines to limit sojourn time.
In order to analyse \tap{} using \textsc{Uppaal}, 
we use the translation given in~\cite{DBLP:conf/formats/Gomez09}. % TODO {Maybe a bit too low-level}
% 
% \todo[inline]{Check whether we need this actually:
% For the remainder of this paper, we apply two slight adaptations related to inputs and the properties discussed above. 
% Whenever inputs and outputs are enabled, we disable inputs implicitly, i.e. we restrict the semantics further
% by adding a transition $(l,\nu) \xrightarrow{i} (l', \nu[r])$ for $i \in \Sigma_I$ only if 
% $\forall l \xrightarrow{g,a,r} l': a \in \Sigma_O \rightarrow \nu \cancel{\models} g$. This follows the reasoning that
% we cannot block outputs and that outputs occur urgently. % Hence, we ensure \emph{isolated outputs} partially on the level of semantics.
% Furthermore, \ldots always enable inputs, i.e. change also the following two sentences }
We implicitly add self-loops to all states $s = (l,\nu)$ for inputs $i$ undefined in $s$, 
% unless outputs are enabled, 
i.e. we add
$(l,\nu) \xrightarrow{i} (l, \nu)$ if $\nu \cancel{\models} \bigvee_{\exists l',r : l \xrightarrow{g,i,r} l'} g$. %and $\forall o \in \Sigma_O: s \centernot{\xrightarrow{o}}$. 
This ensures input enabledness while avoiding \tap{} cluttered with input self-loops. 

The assumptions placed on \acp{SUT} ensure testability~\cite{DBLP:conf/fates/HesselLNPS03}. Assuming 
that \acp{SUT} can be modelled in some modelling formalism is usually referred to as \emph{testing hypothesis}.
Placing the same assumptions on learned models simplifies checking conformance between model and \ac{SUT}. 
The execution of a test sequence on such a model uniquely determines a response \cite{DBLP:journals/tcs/SpringintveldVD01},
and due to input-enabledness we may execute any test sequence. This
allows us to use equivalence as conformance relation between learned models and \ac{SUT}. 
What is more, we can approximate checking equivalence between the learned models and the \ac{SUT} by executing test sequences
on the models and check for equivalence between the \ac{SUT}'s responses and the response predicted by the models. %\todo{maybe remove this to a discussion section}

\subsection{Genetic Programming}
Genetic programming~\cite{DBLP:books/daglib/0070933} is a search-based technique to automatically generate 
programs exhibiting some desired behaviour. Like \emph{Genetic Algorithms}~\cite{DBLP:books/daglib/0019083}, 
it is inspired by nature. Programs,
also called individuals, are iteratively refined by: (1) fitness-based selection followed by (2)
operations altering program structure, like mutation and crossover. Fitness measures are problem-specific 
and may for instance be based on tests. In this case, one could assign a fitness value proportional to the number
of tests passed by an individual. 
The following basic functioning principle underlies genetic programming.
\begin{enumerate}[label=\textbf{\arabic*.}]    
 \item Randomly create an initial population.
 \item Evaluate the fitness of each individual in the population.
 \item If an acceptable solution has been found or the maximum number of iterations has been performed: \textbf{stop} and output the best individual
 \item Otherwise select an individual based on its fitness and apply one of:
 \begin{description}
  \item [Mutation:] change a part of the individual creating a new individual.
  \item [Crossover:] select another individual according to its fitness and combine both individuals
  to create offspring.
  \item [Reproduction:] copy the individual to create a new equivalent individual.
%   \item [Architecture-altering operations]
 \end{description}
\item Form a new population from the newly created individuals and go to Step \textbf{2}.
\end{enumerate}

% Basically, we iteratively refine randomly generated candidate solutions 
% via fitness-based selection followed by mutation, crossover or reproduction.
% Once the search has found a solution with adequate fitness, we stop and output
% the best individual. 
%There are several variations of and additions to this general approach and we will discuss additional details
%we apply in Sect.~\ref{sec:gen_prog} in the following.
Several variations of and additions to this general approach exist
and, in the following, we are going to discuss additional details that
we will apply in Sect.~\ref{sec:gen_prog}.

%\paragraph{Mutation Strength and Parameter Adaptation.}
\emph{Mutation Strength and Parameter Adaptation.}
In meta-heuristic search techniques, like evolution strategies~\cite{DBLP:journals/nc/BeyerS02}, 
mutation strength typically describes the level of change caused by mutations. This parameter
heavily influences search and therefore there exist various schemes to adapt it. 
Since the optimal value for it is problem-specific, it makes sense to evolve throughout 
the search together with the actual individuals. 
Put differently, individuals are equipped with their own mutation strength,
which is mutated during the search. Individuals with good mutation strength
are assumed to survive selection and reproduce. 

%\paragraph{Elitism.}
\emph{Elitism.}
Elitist strategies keep track of a portion of the fittest individuals found so far 
and copy them into the next generation~\cite{DBLP:books/daglib/0019083}.
%Applying such a strate 
This may improve performance, as correctly identified 
partial solutions cannot be forgotten. 

%\paragraph{Subpopulations and Migration.}
\emph{Subpopulations and Migration.}
Due to their nature, genetic algorithms and genetic programming lend themselves to parallelisation.
Several populations may, e.g., be evolved in parallel, which is particularly useful if speciation is applied~\cite{DBLP:conf/kes/NowostawskiP99}.
In speciation, different subpopulations explore different parts of the search space. 
In order to exchange information between the subpopulations, it is common that individuals migrate
between them. 

\section{Genetic Programming for Timed Automata}
\label{sec:gen_prog}
% \subsection{Overview}

\begin{figure}[t]
\centering
% \subfloat[.]{
 \begin{tikzpicture} [->,>=stealth',font=\scriptsize\sffamily,text width = 1.5cm,text centered, minimum height = 0.5cm, node distance = 0.7cm, thick]%
  \node[rounded corners, draw,text width = 1.1cm](testing){Test SUT};
  
  \node[rounded corners, draw, right = 1.2cm  of testing](eval){Evaluate};
  \node[rounded corners, draw, right = 1.2cm  of eval](eval_failing){Evaluate};
  
  \node[trapezium,trapezium left angle=100, trapezium right angle=80,
   draw, above = 0.45cm of eval,text width=1.7cm, minimum height=0.75cm](initial_random){Random Global Population};
  \node[trapezium,trapezium left angle=100, trapezium right angle=80,
   draw, above = 0.45cm of eval_failing,text width=1.7cm, minimum height=0.75cm](initial_random_failing){Random Local Population};
  \node[trapezium,trapezium left angle=100, trapezium right angle=80,
   draw, above right = 0.45cm and -0.9cm of testing,text width=0.5cm, minimum height=0.75cm](sut){SUT};

  \node[diamond, aspect=2.5, draw, below = 0.4 cm of eval,inner sep=-1pt, minimum width = 6em,text width =1.5cm](stop){Stop?};
  \node[rounded corners, draw, below = 0.4cm  of stop](new_pop){Create New Population};
  \node[rounded corners, draw, below left = 0.4cm and -1.4cm of eval_failing](new_pop_failing){Create New Population};
  \node[rounded corners, draw, minimum height]at(eval_failing|-new_pop)(migrate){Migrate};

  \node[trapezium,trapezium left angle=100, trapezium right angle=80,
  trapezium stretches body, draw, below left = 0.2 cm and 0.4cm of new_pop, text width=1.5cm, minimum width=1.5cm](output){Output Fittest};

%   \node[below right = 1.1cm and -1.8cm of new_population_text,inner sep=0pt,outer sep=0pt, align = right,text width = 1.7cm, minimum height=0pt]{$\npop$ \\ times};
 
  \draw[->](testing) --node[above right=0.15cm and -0.4cm,text width=width("traces")]{$\ntest$ traces} (eval);
  \draw[->](initial_random) --node[right,text width=]{$\npop$} (eval);
  \draw[->](initial_random_failing) --node[right,text width=]{$\npop$} (eval_failing);
  
  \draw[->](eval.south) -- (stop.north);
  \draw[->](sut.south-|testing.north) -- (testing.north);
  \draw[->](eval_failing) -- (new_pop_failing);
  \draw[->](eval_failing.340) -- (migrate.20);
  \draw[->](migrate) --node[text width=,above right=0.2cm and -0.4cm]{$\leq \nmig$} (new_pop);
  \draw[->](stop.west) -|node[below right=0cm and 0.2cm,text width=]{yes} (output);
  \draw[->](stop.south) --node[right,text width=]{no} (new_pop);
  \draw[->](new_pop.west) to[out=150, in=210]node[above right=0.2cm and -1cm,text width=]{$\leq \npop+\nmig$} (eval.west);
  \draw[->](new_pop_failing.west) to[out=150, in=210]node[right,text width=]{$\npop$} (eval_failing.west); 
  
  \draw[->](eval.east) --node[above right = 0.15cm and -0.4cm,text width=]{$\mathcal{T}_\mathrm{fail}$} (eval_failing); 
  
\end{tikzpicture}
\label{fig:overview_steps}
\caption{Overview of genetic programming.}
\end{figure}

\begin{figure}[t]
\centering
\begin{tikzpicture}[->,>=stealth',font=\scriptsize\sffamily,text width = 1.3cm,text centered, minimum height = 0.5cm, node distance = 0.7cm, thick]
 \node[diamond, aspect=2, draw, inner sep=-1pt, minimum width = 1em](mutate_choice){Choose Operation};
  
  \node[rounded corners, draw, below = 0.3cm  of mutate_choice](crossover){\scriptsize Crossover};
  \node[rounded corners, draw, left = 0.2cm  of crossover](mutate){\scriptsize Mutate};
  \node[rounded corners, draw, right = 0.2cm  of crossover](inter_crossover){\scriptsize Migration Crossover};
   \node[rounded corners, draw,below = 0.2 of crossover, text width = 1.5cm](select){\scriptsize Fitness-Based Selection};
   \node[rounded corners, draw,below right = 0cm and -0.2cm of select, text width = 1.5cm](apply){\scriptsize Apply Operation};
  
  \node[above= 0.5cm of mutate_choice,inner sep=0pt,outer sep=0pt, text width = 4.5cm, minimum height=0pt, align=center](new_population_text)
  {Create New Population: do $\npop$ times};

  \node[trapezium,trapezium left angle=100, trapezium right angle=80,
 draw, below left = 1cm and 0.1cm of select](pop){Global Population};

%   \node[rounded corners, draw, text width =,below = 0.2cm of pop](migrate){Migrate};
  
  \node[trapezium,trapezium left angle=100, trapezium right angle=80,
   draw, below = 0.3cm of pop](pop_failing){Local Population};
  \node[trapezium,trapezium left angle=100, trapezium right angle=80,
  draw, text width = 1.5cm]at(pop_failing-|inter_crossover)(new_pop){New Global Population};

  \draw[->](mutate_choice) --node[above right=-0.2cm and 0cm,text width=]{$\frac{\pcross}{2}$} (crossover);
 \draw[->](mutate_choice.east) -|node[below right=0cm and -0.1cm,text width=]{$\frac{\pcross}{2}$} (inter_crossover.north);
 \draw[->](mutate_choice.west) -|node[below right=0.1cm and 0cm,text width=]{$1-\pcross$} (mutate.north);
 \draw[->](crossover.south) -- (select);
 \draw[->](mutate.south) -- (select);
 \draw[->](inter_crossover.south) -- (select);
 
 \draw[->](pop.north) |- (select.west);
 \draw[->](pop_failing.east) -|node[text width=](local_to_select){} (select.240);
 \draw[->](select) -| (apply);
 
%  \draw[->](migrate) -- (pop);
 \node[above = 0.5cm of mutate_choice.north,text width=](start_help){};
 \draw[->](start_help) -- (mutate_choice.north);

   \node [text width=,fit = (apply)(select) (mutate)(new_population_text) (crossover) (inter_crossover),
   rounded corners, thick,inner sep=1pt,draw, rounded corners] (fit1) {};
%    \node [fit = (pop) (local_to_select),
%    rounded corners, thick, inner sep=1pt] (fit2) {};
%    \draw[-] (fit2.north east) to [rounded corners] (fit1.south |- fit1.south east) to[rounded corners] 
%    (fit1.south east) to[rounded corners] (fit1.north east) to[rounded corners] 
%    (fit1.north west) to[rounded corners](fit2.north -|fit1.west) -- cycle; %(fit1.south |- fit1.south east);

 \draw[->](new_pop.north|-fit1.south) --node[text width=,right]{$\npop$} (new_pop.north);
 
\end{tikzpicture}
\caption{Creating a new global population.}
\label{fig:new_global_pop}
\end{figure} %\todo{create more accurate figure}

We introduced genetic programming above and discuss
our implementation in the following. 
Figure~\ref{fig:overview_steps} provides an overview of the steps we perform, while
Figure~\ref{fig:new_global_pop} shows the creation of a new population in more detail.

We start with testing of the \ac{SUT}. For that, we generate $\ntest$
test sequences and execute them, to collect $\ntest$ timed traces. Our goal is then 
to genetically program a \ta{} consistent with the collected timed traces. Put differently, we want 
to generate a \ta{} that produces the same outputs as the \ac{SUT} in response to the inputs of the test sequences.
For the following discussion, we say that a \ta{} passes a timed trace $t$
if it produces the same outputs as the \ac{SUT} when simulating the test sequence corresponding to $t$.
Otherwise it fails $t$.

Generally, we evolve two populations of \tap{} simultaneously, a global population which is evaluated on all the traces and a local population which is only evaluated on the traces that fail on the fittest automaton of the global population. 
Both are initially created in the same way and contain $\npop$ \tap.
After initial creation, the global population is evaluated on all $\ntest$ traces.
During that, we basically test the \tap{} and check how many traces each \ta{} passes and assign fitness values accordingly,
i.e., the more passing traces the fitter. Additionally, we add a fitness penalty for model size. 
The local population is evaluated only on a subset $\mathcal{T}_\mathrm{fail}$ of the traces. 
This subset $\mathcal{T}_\mathrm{fail}$ contains all traces which the fittest \ta{} fails, and which likely most of the 
other \tap{} fail as well. With the local population, we are able to explore new parts of the search space more easily
since we may ignore functionality already modelled by the global population. 
We integrate functionality found via this local search into the
global population through migration and migration combined with crossover.
To avoid overfitting to a low number of traces, we ensure that $\mathcal{T}_\mathrm{fail}$ contains at least
$\frac{\ntest}{100}$ traces. If there are less actually failing traces, we add randomly chosen traces from all
$\ntest$ traces to $\mathcal{T}_\mathrm{fail}$.

After evaluation, we stop if we either reached the maximum number of generations $\gmax$, or the fittest \ta{} 
passes all traces and has not changed in $\gchange$ generations. Note that two \ta{} passing all traces may 
have different fitness values depending on model size, i.e. $\gchange$ controls how long 
we try to decrease the size of the fittest \ta{}. The rationale behind this is that \tap{}
of smaller size are less complex and therefore simpler to comprehend. 

If not stopped, we create new populations of \ta, which works slightly differently
for the local and the global population. Figure~\ref{fig:new_global_pop} illustrates the creation of a new global population.
Before creating new \tap{}, existing \tap{} may migrate from the local to the global population. For that, we check each of the fittest $\nmig$ local \ta{} 
and add it to the global population if it passes at least one trace from $\mathcal{T}_\mathrm{fail}$. We generally
set $\nmig$ to $\frac{5\npop}{100}$, i.e. the top five per cent of the local population are allowed to migrate.
After migration, we create $\npop$ new \ta{} through the application of one of three operations:
\begin{itemize}
 \item{\sl with probability $1-\pcross$: } select a \ta{} from the global population and mutate~it
 \item{\sl with probability $\frac{\pcross}{2}$: } select two \ta{} from the global population and perform crossover with these
 \item{\sl with probability $\frac{\pcross}{2}$: } select one \ta{} from each population and perform crossover with these
\end{itemize}
The rationale behind migration combined with crossover is that migrated \tap{} may have
low fitness from a global point of view and will therefore not survive selection. They may,
however, have desirable features which can be transferred via crossover.
For the local population, we perform the same steps, but without any
migration, in order to keep the local search independent.
% hence the probability for crossover between two local \tap{} is $\frac{\pcross}{2}$. 
%We do not migrate from the global to the local population, 
%keeping the local search independent.
% since we do not want influence the local search. 
Once we have new populations, we start a new generation by evaluating the new \tap{}.

A detail not illustrated in Fig.~\ref{fig:overview_steps} is our implementation of elitism. We always keep track 
of the fittest \ta{} found so far for both populations. In each generation, we add these fit
\ta{} to their respective populations after mutation. 

%\paragraph{Parameters.}
\emph{Parameters.}
Our implementation could be controlled by a large number of parameters.
To ease applicability and to avoid the need for meta-optimisation of parameter settings for 
a particular \ac{SUT}, we fixed as many as possible to constant values. 
The actual values, like $\frac{5\npop}{100}$ for $\nmig$, are motivated by 
experiments. The remaining user-specifiable parameters can usually be set
to default values or chosen based on guidelines. For instance, $\npop$, $\gmax$,
and $\ntest$ may be chosen as large as possible, given 
available memory and maximum computation time. 
% \todo{guideline for $\npop$ and $\ntest$, $\gmax$ and $\gchange$}

%TODO: {Add in discussion: Migration without mostly affects early rounds, i.e. improves search speed then and 
%migration with crossover is necessary because it is unlikely that a local \ta{} is globally \ta{} but we may 
% enhance global \ta{} with local functionality through crossover.}

\subsection{Creation of Initial Random Population}
\begin{table}[t]
\centering
\caption{Parameters for Initial Creation of \ta{}}
% \vspace{-0.2cm}
 \begin{tabular}{r|l}
 Name & Short description \\ \Xhline{4\arrayrulewidth} 
 $\Sigma_I$ \&  $\Sigma_O$ & the input and output action labels on edges \\ \hline
 $\nclock$ & number of clocks in the set of clocks $\mathcal{C}$ \\ \hline
 $\cmax$ & approximate largest constant in clock constraints \\ 
 \end{tabular}
\label{tab:parameters}
% \vspace{-0.3cm}
\end{table}

As discussed, we initially create $\npop$ random \tap{}. The
parameters in Table~\ref{tab:parameters} control this creation. Note,
$\cmax$ is an approximation, because mutations may increase constants.
Each \ta{} has initially only two locations, as we intend to increase size and thereby complexity only through
mutation and crossover. Moreover, it is assigned the given action labels and has a set of $\nclock$ clocks. 
During creation, we add random edges, such that at least one edge connects the initial
location to the other location.
% In addition to that, we choose the number of remaining edges probabilistically.

We create edges entirely randomly, whereby the number of constraints in guards as well as the number of 
clock resets are geometrically distributed with fixed parameters. 
The label of an edge, the relational operators and constants in constraints are chosen
uniformly at random from the respective sets $\Sigma$, $\{<,\leq,\geq,>\}$, and $[0 \twodots \cmax]$ 
(operators for outputs exclude~$>$). 
The source and target locations are also chosen 
uniformly at random from the set of locations, i.e. initially we choose from two locations.

If the required number of clocks is not known a priori, we suggest setting $\nclock = 1$
and increasing it only if it is not possible to find a valid \ta{}. A similar approach could be used
for $\cmax$, i.e. setting it to a low value for an initial search.

\subsection{Fitness Evaluation}
\label{sec:fitness}

\subsubsection{Simulation}
We simulate the \tap{} to evaluate their fitness. In the beginning of
this section, we discussed failing and passing trace, but evaluation
is more fine grained.
We execute the inputs of each timed trace and observe produced 
outputs until (1) the simulation is complete, (2) an expected output
is not observed, or (3) output isolation is violated (output non-determinism).
%we expect an ouput, but do not observe it, or 
%(3) we see non-determinism with respect to outputs.

In general, if $\mathcal{T}$ is a deterministic, input-enabled \ta{} with isolated and urgent outputs 
and $\mathit{ts}$ is  a test sequence, then executing $\mathit{ts}$  on $\mathcal{T}$ uniquely determines
a timed trace $\mathit{tt}$. By the testing hypothesis, the \ac{SUT} fulfils these assumptions. 
However, \tap{} generated through mutation and crossover are 
%solely implicitly 
input-enabled, but may show non-deterministic 
behaviour. Hence, simulating a test sequence or a timed trace on a generated \ta{} may follow multiple paths of states.
Some of these paths may produce the expected outputs and some may not. 
Our goal is to find a \tap{} that is both correct, i.e. produces the same outputs as the \ac{SUT},
%and behaves deterministically. 
and is deterministic.
Consequently, we reward these properties with positive fitness.

The simulation function $\textsc{sim}(\mathcal{G}, \mathit{tt})$, simulates a timed trace $\mathit{tt}$
on a generated \ta{} $\mathcal{G}$ and returns a set of timed traces. 
%It builds the basis for fitness computation. 
%This function 
It uses the \ac{TTS} semantics for \ta{} but does not treat outputs as urgent outputs. From 
the initial state $(l_0, \mathbf{0})$, 
where $l_0$ is the initial location of $\mathcal{G}$, 
it performs the following steps for each $t_i e_i \in \mathit{tt}$ with $t_0 = 0$:

\begin{enumerate}[label=\arabic*.]
 \item From state $q = (l, \nu)$
 \item Delay for $d=t_i - t_{i-1}$ to reach $q^d=(l,\nu+d)$
 \item If $e_i \in \Sigma_I$, i.e. it is an input:
 \begin{enumerate}[label=3.\arabic*.]
  \item If $\exists o \in \Sigma_O, d^o \leq d: (l,\nu+d^o) \xrightarrow{o}$, 
  i.e. an output would have been possible while delaying or at time $t_i$
  \item[$\rightarrow$] then mark $e_i$ 
  \item If $\exists q^1,q^2, q^1\neq q^2 : q^d \xrightarrow{e_i} q^1 \land q^d \xrightarrow{e_i} q^2$
  \item[$\rightarrow$] then mark $e_i$ 
  \item For all $q'$ such that $q^d \xrightarrow{e_i} q'$
  \item[$\rightarrow$] carry on exploration with $q'$
 \end{enumerate}
\item If $e_i \in \Sigma_O$, i.e. it is an output:
 \begin{enumerate}[label=4.\arabic*.]
  \item If $\exists o \in \Sigma_O, d^o < d: (l,\nu+d^o) \xrightarrow{o}$, 
  i.e. an output would have been possible while delaying
  \item[$\rightarrow$] stop exploration
  \item If $\exists q^1,q^2, q^1\neq q^2 : q^d \xrightarrow{e_i} q^1 \land q^d \xrightarrow{e_i} q^2$ or 
  $\exists o, o\neq e_i : q^d \xrightarrow{o}$
  \item[$\rightarrow$] stop exploration
  \item If there is a $q'$ such that $q^d \xrightarrow{e_i} q'$
  \item[$\rightarrow$] carry on exploration with $q'$
 \end{enumerate}
\end{enumerate}
The procedure shown above allows for two types of non-determinism. During delays before
executing an input, we may ignore outputs (3.1.) and we may explore multiple
paths with inputs (3.3.). We mark these inputs to be non-deterministic, through (3.1.\ and 3.2.).
Since we explore multiple paths, a single input $e_i$ may be marked along one path but not marked along
another path. In contrast, we do not allow for non-determinism with respect to outputs
to avoid issues with trivial \tap{} which produce each output all the time. These would completely
simulate all traces, but would not be useful. 

During exploration, $\textsc{sim}(\mathcal{G}, \mathit{tt})$ collects
and returns timed traces $tts$, which 
are basically prefixes of $\mathit{tt}$ but with marked and unmarked inputs. 
For fitness computation, we defined four auxiliary functions. The first one assigns 
a simulation verdict, which is $\pass$ if $\mathcal{G}$ behaves deterministically and produces the expected
outputs. It is $\nondet$ if it produces the correct outputs along at least one execution path, but behaves 
non-deterministically. Otherwise it is $\fail$.
\begin{align*}
%\text{Let } \mathit{tts} &= \textsc{sim}(\mathcal{G}, \mathit{tt}) \text{ be collected timed traces} \\
 \textsc{verdict}&(tts) = \\ &\begin{cases}
                \pass &\text{ if } |\mathit{tts}| = 1 \land \mathit{tt} \in \mathit{tts} \\
                \nondet &\text{ if }|\mathit{tts}| > 1 \land \exists tt' \in \mathit{tts}: |tt'| = |tt| \\
                \fail &\text{ otherwise }  \\
               \end{cases}
\end{align*}
%
%The
Function $\textsc{steps}(\mathit{tts})$ returns the maximum number
of unmarked inputs in a trace in $\mathit{tts}$, i.e. the deterministic steps, and 
$\textsc{out}(\mathit{tts})$ returns the number of outputs along the longest traces in $\mathit{tts}$.
Finally, $\textsc{size}(\mathcal{G})$ returns the number of edges. %of $\mathcal{G}$.

Note that a \ta{}, which produces a $\pass$ verdict for all timed traces, behaves
equivalently to the \ac{SUT} for these traces. 
As indicated at the end of Sect.~\ref{sec:ta_prelim},
we approximate equivalence testing via simulation with $\textsc{sim}$. 
%, also if we interpret
%that the \ta{} produces outputs urgently. 

\subsubsection{Fitness Computation}
In order to compute the fitness, we assign the weights $w_{\pass}$, $w_{\nondet}$, $w_{\fail}$, $w_{\textsc{steps}}$, 
$w_{\textsc{out}}$, and $w_{\textsc{size}}$ to the gathered
information of $\mathcal{G}$.
Basically, we give some positive fitness for deterministic steps, 
correctly produced outputs, and verdicts, but penalise size.
Let $\mathcal{TT}$ be the timed traces on which $\mathcal{G}$ is evaluated. The fitness
$\textsc{fit}(\mathcal{G})$  with  $tts =\ \textsc{sim}(\mathcal{G},tt)$  is then:
\begin{align*}
 \textsc{fit}(\mathcal{G}) =\ &\sum_{tt\in \mathcal{TT}} \textsc{fit}(\mathcal{G},tt) - w_{\textsc{size}}\ \textsc{size}(\mathcal{G})\; \text{ where } \\
 \textsc{fit}(\mathcal{G},tt) =\ &w_{\textsc{verdict}(tts)} +
 w_{\textsc{steps}}\ \textsc{steps}(tts) \\ &+ w_{\textsc{out}}\ \textsc{out}(tts)
% & \text{ with } tts =\ \textsc{sim}(\mathcal{G},tt) \\
\end{align*}
Fitness evaluation adds further parameters. We identified guidelines 
for choosing them adequately. We generally set $w_{\fail} = 0$ and use $w_{\out}$ as basis for other weights.
Usually, we set $w_{\step} = w_{\out}/2$ and $w_{\pass} = k \cdot l \cdot w_{\out}$, where $l$ is the average length of test sequences
and $k$ is a small natural number, e.g. $4$. More important than the exact value of $k$ is setting $w_{\nondet} = w_{\pass} / 2$
which gives positive fitness to correctly produced timed traces but with a bias towards deterministic solutions.
% The weight $w(\textsc{size})$ for the penalty term for automaton size should take the number of tests into account and 
% should be low if a low number of tests is available. If, e.g., an output is observed in only a few tests then adding 
% an edge required for that output may decrease fitness if $w(\mathit{size})$ is too large. In doubt, it may be set to, e.g., $w(\step)$.
The weight $w_{\textsc{size}}$ should be chosen low, such that it does not prevent adding of necessary 
edges. We usually set it to $w_{\step}$. 
It needs to be non-zero, though. Otherwise an acceptable solution could be a tree-shaped automaton 
exactly representing $\mathcal{TT}$ without generalisation. 

\subsection{Creation of New Population}
We discussed how we create a new population at the beginning
of this section on the basis of Fig.~\ref{fig:new_global_pop} and we will now present
details of the involved steps. 

%\paragraph{Migration.}
\subsubsection{Migration}
%Initially, we distinguished only passed and failed traces, but 
%later introduced a third verdict for non-deterministically passed traces.
In the context of migration, we consider non-deterministically passed traces to be failed,
i.e. $\mathcal{T}_\mathrm{fail}$ contains all traces for which 
the fittest \ta{} of the global population produces a verdict other than $\pass$.
The rationale behind this is that we want to improve for 
traces with both $\fail$ and $\nondet$ verdict. 

%\paragraph{Selection.}
\subsubsection{Selection}
We use the same selection strategy for mutation and crossover, except that 
crossover must not select the same parent twice. In particular,
%we apply a combination of \emph{truncation} and \emph{probabilistic tournament} selection.
we combine \emph{truncation} and \emph{probabilistic tournament} selection:
first, we discard the $(\npop-\nsel)$ worst-performing non-migrated \tap{} (truncation).
Then, for each individual selection, we perform a tournament selection
from the remaining $\nsel$ \tap{} of the global population and migrated \tap{}. 
Probabilistic tournament selection \cite{DBLP:conf/cec/HingeeH08} randomly chooses a set of $\ntourn$ 
\tap{} and orders them by their fitness. It then selects 
the $i^{th}$ \ta{} with probability $p_i$, which we set to $p_i = \ptourn (1-\ptourn)^{i-1}$ for 
$i \in [1\twodots \ntourn-1]$ and $p_{\ntourn} = (1-\ptourn)^{\ntourn-1}$,
with $\ntourn = 10$ and $\ptourn = 0.5$.

Truncation selection is mainly motivated by the observation that it increases convergence
speed during early generations by concentrating on the fittest \tap{}. 
However, it can be expected to cause a larger loss of diversity
than other selection mechanisms~\cite{DBLP:journals/ec/BlickleT96}. As a result, search may converge 
to a suboptimal solution, because \tap{} that might need several generations to evolve to an 
optimal solution are simply discarded through truncation. Therefore, we gradually increase
$\nsel$ until it becomes as large as $\npop$ such that no truncation is applied in later generations.
For the same reason, we do not discard migrated \tap{}, since they may
possess valuable features. 

\begin{table}[t]
\centering
\caption{Mutation Operators}
% \vspace{-0.2cm}
 \begin{tabular}{l|p{6cm}}
 Name & Short description \\ \Xhline{4\arrayrulewidth} 
 add constraint & adds a guard constraint to an edge \\ \hline
 change guard & select edge and create a random guard if the edge does not have a guard, otherwise mutate a constraint of its guard \\ \hline
 change target & changes the target location of an edge \\ \hline
 remove guard & remove either all or a single guard constraint from an edge\\ \hline
 change resets & remove or add clocks to the clock resets of an edge \\ \hline
 remove edge & removes a selected edge \\ \hline 
 add edge & adds an edge connecting randomly chosen locations \\ \hline
 sink location &  adds a new location \\ 
  \Xhline{4\arrayrulewidth}
  merge location & merges two locations \\ \hline
  split location & splits a location $l$ by creating a new location $l'$ and redirecting an edge reaching $l$ to $l'$ \\ 
  \Xhline{4\arrayrulewidth}
  add location & adds a new location and two edges connecting the new location to existing locations \\ \hline
  split edge & replaces an edge $e$ with either the sequence $e' \cdot e$ or $e \cdot e'$ where $e'$ is a new random edge (adds a location to connect $e$ and $e'$)\\ 
 \end{tabular}
\label{tab:operators}
% \vspace{-0.3cm}
\end{table}

%\paragraph{Application of Mutation Operators.}
\subsubsection{Application of Mutation Operators}
We implemented mutation operators for changing all aspects of \tap{}, such as adding and removing clock constraints. 
Table \ref{tab:operators} lists all operators. % together with a short description.
Whenever an operator selects an edge or a location, the selection is random, but favours locations and edges which are associated
with faults and non-deterministic behaviour. We augment \tap{} with such information during fitness evaluation.
To create an edge, we create random guards and reset sets, and choose a random label, like for the initial creation of 
\tap{}. 

The mutation operators form three groups separated by bold horizontal lines. The first and largest group contains basic operators, which 
are sufficient to create all possible automata. The second group is motivated by the basic principle behind automata learning
algorithms. Passive algorithms often start with a tree-shaped representation of traces and transform this representation
into an automaton via iterated state-merging~\cite{delaHiguera_2010}. Active learning algorithms on the other hand usually 
start with a low number of locations and add new locations if necessary. This can be interpreted as splitting of existing locations,
an intuition which also served as a basis for test-case generation in active automata learning~\cite{DBLP:conf/nfm/AichernigT17}.
The last two operators are motivated by observations during experiments: \emph{add location} increases the automaton size
but avoids creating deadlock states, unlike the operator \emph{sink location}. \emph{Split edge} addresses issues related to input enabledness,
where an input $i$ is implicitly accepted without changing state, although an edge labelled $i$ should change the state. The operator
aims to introduce such edges.
For mutation, we generally select one of the operators uniformly at random.

%\paragraph{Simplification.}
\subsubsection{Simplification}
In addition to mutation, we apply a simplification procedure. % for \tap{}. 
It changes the syntactic representation of \tap{} without affecting semantics, by, e.g., removing
unreachable locations and self-loops for inputs which do not reset clocks. This limits the search to relevant
parts of the search space, i.e. we do not mutate unreachable edges. The parameter $\gsimp$ 
specifies the number of generations between executions of simplification. 
Note that we check
only the underlying graph of the \ta{}, but do not consider clock values to ensure fast operation.

%\paragraph{Crossover.}
\begin{figure}[t]
 \begin{algorithmic}[1]
% \footnotesize
 \State $(l^1,l^2) \gets \text{current product location}$
 \ForAll{$l^1 \xrightarrow{g^1,a,r^1} {l^1}'$}
 \If{$l^2 \xrightarrow{g^2,a,r^2} {l^2}'$} \Comment{synchronise on
   label $a$}
 \State $\textsc{add}$ ($(l^1,l^2) \xrightarrow{\textsc{choose}(g^1,g^2),a,\textsc{choose}(r^1,r^2)} ({l^1}',{l^2}')$
 \Else
 \State $\textsc{add}$ ($(l^1,l^2) \xrightarrow{g^1,a,r^1} ({l^1}',\textsc{chooseFrom}(L^2))$
 \EndIf
 \EndFor
 \ForAll{$l^2 \xrightarrow{g^2,a,r^2} {l^2}'$ s.t. $\nexists g^1,r^1,{l^1}':l^1 \xrightarrow{g^1,a,r^1} {l^1}'$}
 \State $\textsc{add}$ ($(l^1,l^2) \xrightarrow{g^2,a,r^2} (\textsc{chooseFrom}(L^1),{l_2}')$
 \EndFor
\end{algorithmic}
 \caption{Crossover for location $(l^1,l^2)$.}
 \label{alg:crossover}
\end{figure}

\subsubsection{Crossover}
We basically implement crossover as a randomised product of two parents. Briefly, it works as follows. 
Let $L^1$ and $L^2$
be the locations of the two parents and let ${l_0}^1$ and ${l_0}^2$ be their respective initial locations, 
then the locations of the offspring are given by $L^1 \times L^2$.
Beginning from ${l_0}^1$ and ${l_0}^2$ and the initial product location $({l_0}^1,{l_0}^2)$, we explore 
both parents in a breadth-first manner and add edges via the algorithm shown in \figurename{}~\ref{alg:crossover}.
Crossover synchronises on action labels and adds edges common to both parents, while randomly choosing the guard and 
resets from one of the parents. Edges present in only one parent (Line $5-8$) are added as well,
but the target location for the other parent is chosen randomly. 
The random auxiliary function $\textsc{choose}(a,b)$ returns either $a$ or $b$ with equal probability and
$\textsc{chooseFrom}(L)$ chooses a location in $L$ uniformly.

To avoid creating excessively large offspring, we stop the exploration and consequently adding edges, once
the number of reachable product locations is equal to $\max(L^1,L^2)$, i.e. the offspring may not 
have more locations than both parents. The reachability check only considers the graph underlying the \ta{}
and ignores guards due to efficiency reasons.

%\paragraph{Mutation Strength.}
\subsubsection{Mutation Strength}
To control mutation strength, we augment each \ta{} with a probability $\pmut$. 
Basically, we perform iterated mutation and stop with $\pmut$ after each mutation.
% Offspring created through mutation are either assigned the parent's $\pmut$, 
% $\pmut$ increased by multiplication with $\frac{10}{9}$, or $\pmut$ decreased
% by multiplication with $\frac{9}{10}$. These changes are constrained to not exceed
% a maximum of $0.9$ or a minimum of $0.1$. 
%
\tap{} created by mutation are assigned the parent's $\pmut$, 
$\pmut$ increased by multiplication with $\frac{10}{9}$, or $\pmut$ decreased
by multiplication with $\frac{9}{10}$. These changes are constrained to not exceed
the range $[0.1,0.9]$. \tap{} created via crossover are assigned the average 
$\pmut$ of both parents.
In the first generation, we set $\pmut$ of all \tap{} 
to the user-specified $\pmutinit$.  The search is insensitive to this parameter as 
it quickly finds suitable values for $\pmut$ via mutation.

\section{Case Studies}
\label{sec:case_studies}
Our evaluation is based on four manually created
and $40$ randomly generated \tap{}, which serve as our \acp{SUT}. Using \tap{} 
provides us with an easy way of checking whether we found the correct model, however,
our approach and our tool are general enough to work on real black-box implementations. 
Our algorithms are implemented in Java. 
%The evaluation was performed with a Java implementation. 
A demonstrator with a GUI %graphical user interface
%to this implementation 
is available in the supplementary material, which also includes 
Graphviz dot-files of the \tap{}~\cite{supp_material}. The demonstrator
allows repeating all experiments presented in the following with freely
configurable parameters. Moreover, the search progress can be inspected
anytime. The user interface lists the fittest \tap{} for each generation
and provides an option to visualise each of them along with the timed 
traces used for learning. 

%\footnote{Supplementary 
%material:~\url{https://figshare.com/s/442ac3717e19b9bbf1c6}}. %  used in the evaluation

% In the evaluation, we used simulated time, but the technique is generally 
% applicable for systems producing timed traces. 
For the evaluation, we generated timed traces by simulating $\ntest$ random test sequences
on the \acp{SUT}. The inputs in the test sequences were selected uniformly at random from the available inputs. 
The lengths of the test sequences
are geometrically distributed with a parameter $\ptest$, which is set to $0.15$ unless otherwise noted.
To avoid trivial timed traces, we ensure that all test sequences cause
at least one output to be produced. The delays in test sequences
were chosen probabilistically in accordance with the user-specified largest constant 
$\cmax$. Additionally one could specify important constants used in the \acp{SUT}, which could be gathered from a requirements document. Specifying
appropriate delays helps to ensure that the \acp{SUT} are covered sufficiently well by the test sequences.

%\paragraph{Measurement Setup and Criteria.}
\emph{Measurement Setup and Criteria.}
The measurements were done on a notebook with 16 GB RAM and an 
Intel Core i7-5600U CPU operating at 2.6 GHz. Our main goal is to show
that we can learn models in a reasonable amount of time, but further improvements
are possible, e.g., via parallelisation. We use a training set and a test set
for evaluation, each containing $\ntest$ timed traces. First, we learn from the training
set until we find a \tap{} which produces a $\pass$ verdict for all traces.
Then, we simulate the traces from the test set and report all
traces leading to a verdict other than $\pass$ as erroneous. Note that since 
we generate the test set traces through testing, there are no negative traces.
In other words, all traces are observable and can be considered positive. Consequently, 
notions like precision and recall do not apply to our setting.

Our four manually created \tap{}, with number of locations and $\cmax$ in parentheses, are called
\ac{CAS} ($14,30$), Train ($6,10$), Light ($5,10$), and \ac{PC} ($26,10$). 
All of them require one clock. %\todo{add references}
The \ac{CAS} for instance served as a benchmark for test-case generation for timed systems~\cite{DBLP:conf/tap/AichernigLN13,DBLP:conf/birthday/AichernigLT16}.
Different versions of the Train and Light \tap{} have been used as examples in real-time verification~\cite{DBLP:conf/sfm/BehrmannDL04} and 
and variants of them are included in the demos distributed with the real-time model-checker \textsc{Uppaal}~\cite{DBLP:journals/sttt/LarsenPY97}
and the real-time testing tool \textsc{Uppaal} \textsc{Tron}~\cite{DBLP:conf/fortest/HesselLMNPS08}. 
Untimed versions of the particle counter (\ac{PC}) were examined in model-based testing~\cite{DBLP:conf/tap/AichernigAJKKSS14,DBLP:journals/entcs/AichernigT16}.

In addition to the manually created timed systems, we have four categories of random \tap{}, each containing ten \tap{}:
C$15/1$, C$20/1$, C$6/2$, C$10/2$, where the first number gives the number of locations and the second the
number of clocks. \tap{} from the first two categories have alphabets containing $5$ distinct inputs and $5$ distinct outputs, while
the \tap{} from the other two categories have $4$ inputs and $4$ outputs. For all random \tap{}, we have $\cmax=15$.

We used similar configurations for all experiments. Following the suggestions in Sect.~\ref{sec:gen_prog}, 
we set the fitness weights
to $w_{\textsc{out}} = 0.25$, $w_{\textsc{steps}} = \frac{w_{\textsc{out}}}{2} = w_{\textsc{size}}$,
$w_{\pass} = \frac{4 w_{\textsc{out}}}{\ptest}$, $w_{\nondet} = \frac{\pass}{2}$, and $w_{\fail}=0$,
with the exception of \ac{CAS}. Since the search frequently got trapped in minima with non-deterministic 
behaviour, we set $w_{\textsc{out}} = \frac{w_{\textsc{steps}}}{2}$, i.e. we valued deterministic steps
more than outputs, and $w_{\nondet} = -0.5$, i.e. we added a small penalty for non-determinism.
Other than that, we set $\gmax=3000$, $\npop = 2000$, the initial $\nsel=\frac{\npop}{10}$, $\ntest = 2000$,
$\pcross = 0.25$, $\gchange = 10$, $\pmutinit =0.33$, and $\gsimp=10$, with the following
exceptions. Train and Light require less effort, thus we set $\npop=500$. The categories
C$10/2$, C$15/1$, and C$20/1$ require more thorough testing, so we configured $\ntest = 4000$ for C$10/2$ and C$15/1$, and 
$\ntest = 6000$ with $\ptest = 0.1$ for C$20/1$.

All learning runs were successful by finding a \tap{} without errors on the training set, except
for two cases, one in C$10/2$ and one in C$20/1$. For the first, we repeated the 
experiment with a larger population $\npop=6000$, resulting in successful learning. For the random \tap{}
in C$20/1$, we observed a similar issue as for \ac{CAS}, i.e. non-determinism was an issue, but used
another solution. In some cases, crossover may introduce non-determinism, thus we decreased the probability for crossover
$\pcross$ to $0.05$ and learned the correct model.
\begin{table*}[t]
\centering
\caption{Measurement Results}
% \vspace{-0.2cm}
 \begin{tabular}{l|c|c|c}
  \tap{} & test set errors & generations & time \\ \Xhline{4\arrayrulewidth}
\ac{CAS} & $0$&$103$ / $293.0$, $299.3$ / $861$&$25.4 m$ / $1.4 h$, $1.8 h$ / $5.9 h$\\ \hline
Train & $0$&$57$ / $90.5$, $94.5$ / $161$&$5.3 m$ / $9.1 m$, $8.6 m$ / $14.6 m$\\ \hline
Light & $0$&$42$ / $77.5$, $84.5$ / $240$&$3.2 m$ / $7.4 m$, $8.7 m$ / $31.1 m$\\ \hline
\ac{PC} & $0$ / $0.0$, $0.4$ / $4$&$423$ / $1187.5$, $1137.3$ / $2745$&$1.2 h$ / $4.0 h$, $4.2 h$ / $11.6 h$\\ \hline
C$15/1$ & $0$ / $2.0$, $1.8$ / $6$&$201$ / $404.5$, $401.3$ / $746$&$1.4 h$ / $3.1 h$, $3.2 h$ / $6.6 h$\\ \hline
C$20/1$ & $0$ / $0.0$, $1.0$ / $6$&$45$ / $451.0$, $665.8$ / $1798$&$23.4 m$ / $6.7 h$, $7.4 h$ / $18.3 h$\\ \hline
C$6/2$ & $0$ / $0.0$, $0.5$ / $3$&$18$ / $68.5$, $176.9$ / $709$&$9.4 m$ / $43.9 m$, $1.8 h$ / $7.6 h$\\ \hline
C$10/2$ & $0$ / $2.5$, $2.6$ / $8$&$73$ / $239.0$, $344.9$ / $984$&$35.8 m$ / $3.1 h$, $3.4 h$ / $9.3 h$\\ 
\end{tabular}
\label{tab:results}
% \vspace{-0.5cm}
\end{table*}
Table \ref{tab:results} shows the learning results. The column \emph{test set error} contains $0$, if there were
no errors on the test set. Otherwise, each cell in the table contains, from left to right, the minimum,
the median and the mean, and the maximum computed over $10$ runs for manually 
created \tap{} and over $10$ runs for each random category, i.e. one run per random~\ta{}.

The test set errors are generally low, so our approach generalises well and does not simply overfit 
to the training data. We also see that manually created systems produced no test set errors except in a single run, 
% while especially the most complex category of the randomly generated systems, C$10/2$, led to test set errors. 
while the more complex, random \tap{} led to errors. 
However, also for them the relative number of errors was at most two thousandths ($8$ errors out of
$4000$ tests). Such errors may, e.g., be caused by slightly too loose or
too strict guards on inputs. We believe that the computation time of at most $18.3$ hours is acceptable, especially
considering that fitness evaluation, as the most time-consuming part, is parallelisable.
Finally, we want to emphasise that we identified parameters which almost consistently produced good results.
In the exceptions where this was not the case, it was simple to adapt the configuration
by observing how the search evolved. 

The size of our \tap{} in terms of number of locations ranges between $5$ and $26$. To model real-world systems, it is therefore
necessary to apply abstraction during the testing phase, which collects timed traces. Since model learning requires thorough 
testing, abstraction is commonly used in this area. Consequently, this requirement is not a strong limitation.
Several applications of automata learning show that implementation flaws can be detected by analysing learned abstract models, e.g.,
in protocol implementations~\cite{DBLP:conf/uss/RuiterP15,DBLP:conf/icst/TapplerAB17,DBLP:conf/cav/Fiterau-Brostean16}.

\begin{figure*}[t]
\centering
\clipbox{0 0 0 0.75cm}{
 \includegraphics[width=.9\textwidth]{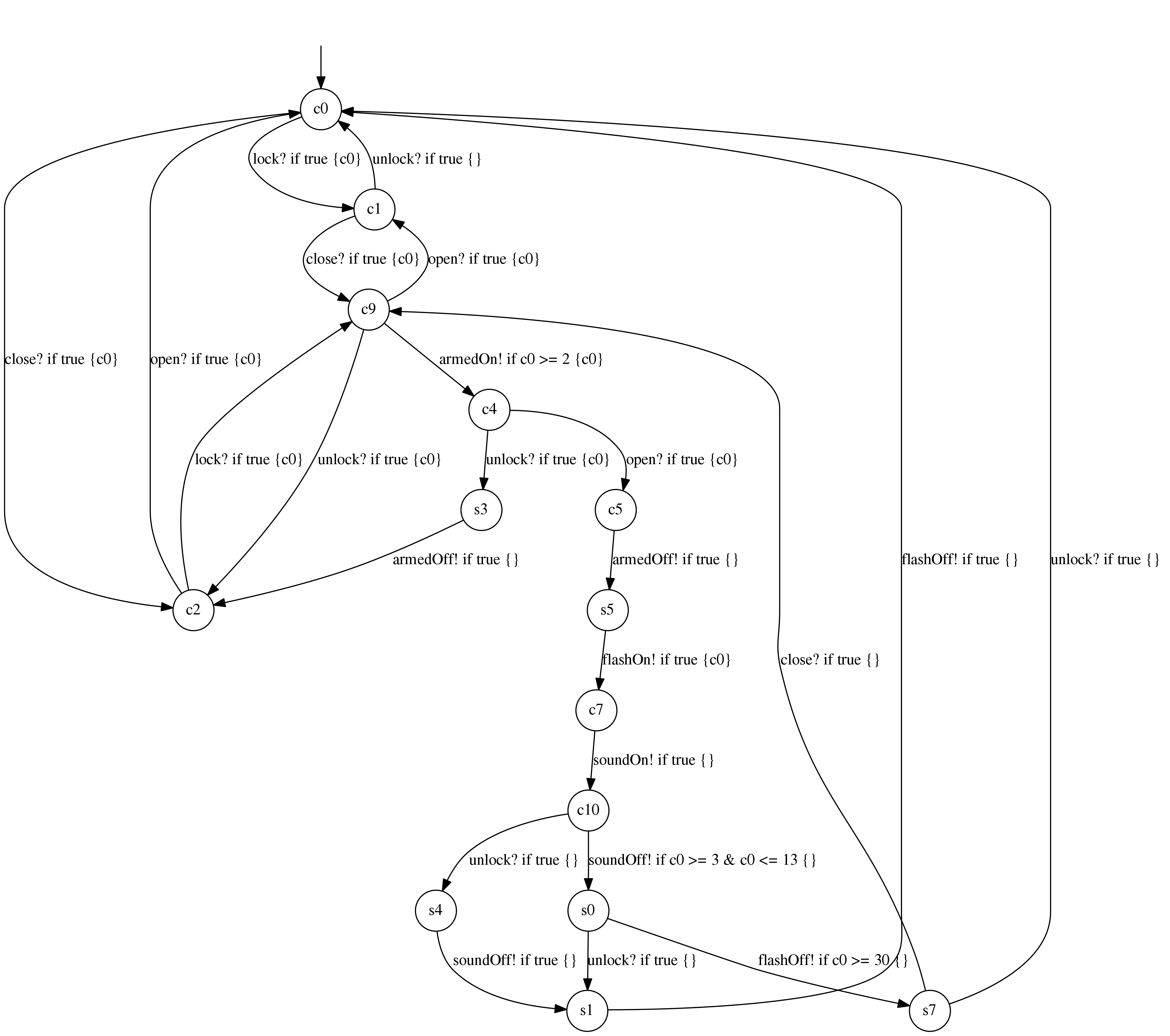}
 }
 \caption{Learned model of the \acf{CAS}.}
 \label{fig:cas}
\end{figure*}

Figure~\ref{fig:cas} shows a learned model of the \ac{CAS} as it is produced by our tool (no post-processing). 
It is observably equivalent to the true system 
generating the timed traces used for learning. There is thus no test sequence which distinguishes
the system from the learned model. Note that the model is also well comprehensible. This is due to 
the fitness penalty for larger systems and due to implicit input-enabledness. Both measures target the 
generation of small models containing only necessary information. The \ac{CAS} model only contains a few unnecessary
clock resets and an ineffective upper bound in the guard of the edge labelled \emph{soundOff!}, but removing any 
edge would alter the observable behaviour.
Therefore, our approach may enable manual inspection of black-box timed systems, which is 
substantially more difficult and labour-intensive given only observed timed traces.

% It is noteworthy that other kinds automata used in automata learning are not able to model the \ac{CAS}.
% The model shown in Fig.~\ref{fig:cas} contains the behaviour: if the car doors are closed (\emph{close?} input)
% and locked (\emph{lock?} input) for $2$ time units, the alarm system should be armed (\emph{Armed!} output).
% Any input 

% \section{Related Work}

\section{Conclusion}
\label{sec:conc}
%\paragraph{Summary.}
\subsection{Summary.}
We presented an approach to learn deterministic \tap{} with urgent
outputs, an important subclass for testing timed systems~\cite{DBLP:conf/fates/HesselLNPS03}.
The learned models may reveal flaws during manual inspection and enable verification of black-box systems via model-checking.
% Genetic programming, a meta-heuristic search-based technique, serves as the basis.
% In our instantiation of this framework, we parallelized search by evolving 
Genetic programming, a meta-heuristic search technique, serves as a basis.
In our implementation of it, we parallelise search by evolving 
two populations simultaneously and developed techniques for mutation, crossover, 
and for a fine-grained fitness-evaluation. 
%We target a subclass of \tap{}, namely deterministic \tap{} with output urgent behaviour, 
%The targeted subclass of \tap{}
%forms the basis for other test-based work for timed systems~\cite{DBLP:conf/fates/HesselLNPS03}.
We evaluated the technique on non-trivial \tap{} with up to $26$
locations. We could learn all 44 \tap{} models, only two random \tap{}
needed a small parameter adjustment. 
%We conclude that the technique shows great potential, especially considering the various 
%possible extensions discussed in the following. These may enhance performance, expressiveness of learned models
%and may broaden the scope of the technique. 

\subsection{Related Work.}
\label{sec:rel_work}
%Closely related work with respect to the domain has been performed by Verwer et al.~\cite{verwer2007algorithm,DBLP:conf/icgi/VerwerWW10}.
%They passively learn (probabilistic) deterministic real-time automata

Verwer et al.~\cite{verwer2007algorithm,DBLP:conf/icgi/VerwerWW10} passively learned real-time automata
via state-merging. 
These \tap{} measure the time between two consecutive events and use guards in the form of intervals, i.e. 
they have a single clock which is reset on every transition. Verwer et
al.\, do not distinguish between inputs and outputs, and hence, do not assume
output urgency. 
Improvements of~\cite{DBLP:conf/icgi/VerwerWW10} were presented in \cite{DBLP:conf/nfm/MediouniNBB17}.
Similarly, Mao et al. applied state-merging for learning continuous 
time Markov chains \cite{DBLP:journals/ml/MaoCJNLN16}.
A state-merging-based learning algorithm for more general stochastic timed systems 
has been proposed by de Matos Pedro et al.~\cite{DBLP:conf/isola/PedroCS12}. They target 
learning generalised semi-Markov processes, which are generated by stochastic timed automata.
All these techniques have in common that they consider timed systems where
the relation between different events is fully described by a system's structure. 
Pastore et al.~\cite{DBLP:conf/icst/PastoreMM17} learn specifications capturing the duration 
of (nested) operations in software systems. A timed trace therefore includes for each operation its start
and its end, i.e. the trace records two related events. Their algorithm \emph{Timed k-Tail}
is based on the passive learning technique \emph{k-Tail}, which they extended to handle timing aspects.

Grinchtein et al.~\cite{DBLP:journals/tcs/GrinchteinJL10,DBLP:conf/concur/GrinchteinJP06} described 
active learning approaches for deterministic event-recording automata, a subclass of \tap{} with one clock 
per action. The clock corresponding to an action is reset upon its execution essentially recording the time since the action has occurred.
While the expressiveness of these automata suffices for many applications, the runtime complexity of the described
techniques is high and may be prohibitive in practice. At the time of writing, there is no implementation to actually measure runtime.
% Furthermore, event-recording automata are not suited for applications which require input enabledness.
% Input edges leaving the state unchanged do not exist in these automata, as edges
% always reset exactly one clock. 
Furthermore, this kind of TA cannot model certain timing patterns,
e.g., in the case of input enabledness where always reseting a clock
may not be appropriate.
Jonsson and Vaandrager~\cite{mealy_timer} also note that the learning approaches presented 
previously~\cite{DBLP:journals/tcs/GrinchteinJL10,DBLP:conf/concur/GrinchteinJP06} are complex 
and developed a more practical active-learning approach for Mealy machines with timers.
Currently, there is no implementation for this approach as well. A further drawback is that
input edges cannot be restricted via guards.

Meta-heuristic search as an alternative to classical model learning such as $L^*$ has, e.g., been proposed 
by Lai et al.~\cite{DBLP:conf/qsic/LaiCJ06} in the context of adaptive model-checking. They 
apply genetic algorithms and assume the number of states to be known. Lucas and Reynolds in contrast compared
state-merging and evolutionary algorithms, but also fixed the number of states for runs of the latter~\cite{DBLP:conf/cec/LucasR03}.

Lefticaru et al.\ similarly assume the number of states to be known and induce state machine models via genetic algorithms~\cite{DBLP:conf/bci/LefticaruIT09}.
Their goal, however, is to synthesise a model satisfying a specification given in temporal logics. 
Early work suggesting such an approach was performed by Johnson~\cite{DBLP:conf/eurogp/Johnson07}, which
like our approach does not require the solution size to be known.
% A similarity to our approach is that Johnson does not require the solution size to be known. 
% Instead, he allows automata to grow via mutation. 
In contrast, Johnson does not apply crossover. %, 
% noting that it is not clear how to carry out such an operation for automata. 
Further synthesis work from Katz and Peled
\cite{DBLP:journals/sttt/KatzP17} tries to infer a correct
program or model on the source code level,  
%but on the level
%of source code. However, 
%these works have a different goal. They 
%they try to synthesise a correct program or model,
while we aim at synthesising a model representing a black-box system. %Errors present in the system should be reflected in the model.

Evolutionary methods have been combined with testing in several areas: Abdessalem et al.~\cite{Abdessalem:2018:TVC:3180155.3180160} use evolutionary algorithms for the generation of test scenarios and learn decision trees to identify critical scenarios. Using the learned trees, they can steer the test generation towards critical scenarios. The tool Evosuite by Fraser and Arcuri~\cite{ESECFSE11} uses genetic operators for optimising whole test suites at once, increasing the overall coverage, while reducing the size of the test suite. Walkinshaw and Fraser presented Test by Committee, test-case generation
using uncertainty sampling~\cite{DBLP:conf/icst/WalkinshawF17}. The approach is independent of the type
of model that is inferred and an adaption of Query By Committee, a technique
commonly used in active learning. In their implementation, they infer several
hypotheses at each stage via genetic programming, generate random tests and
select those tests which lead to the most disagreement between the inferred hypotheses. In contrast to most other works considered, their implementation infers
non-sequential programs. It infers functions mapping from numerical inputs to
single outputs.

The work by Steffen et al.~\cite{1271205,DBLP:conf/sfm/SteffenHM11} is another good showcase for the strong possible relation between testing and model learning. They combine both areas, by performing black box tests and using the results to generate a model. Contrary to our work, they perform active learning, i.e., they use the intermediate versions of the learned models to guide the test generation. For a more comprehensive overview of combinations of learning and testing, we refer to~\cite{DBLP:conf/dagstuhl/AichernigMMTT16}.

\subsection{Future Work.}
%\label{sec:future}
As indicated above, our technique is entirely passive, i.e. we learn from a set of timed traces (test observations),
collected beforehand by random testing.
There is no feedback from genetic programming to testing.
In contrast to this, model-based testing could be applied to 
find discrepancies between the \ac{SUT} and learned models~\cite{DBLP:conf/fm/WalkinshawDG09}.
These may then be used to iteratively refine the models. 
Active testing based on intermediate learned models may improve coverage of the \ac{SUT} 
while requiring fewer tests, since we would benefit from additional knowledge about the system behaviour.
This may therefore lead to improved accuracy of the model and increased
performance through a reduction of tests and testing time. 
We are currently investigating possible implementations of active learning. 

% Based on these, we learn a model of the \ac{SUT} without interacting 
% with the \ac{SUT}. Since we initially have very limited information, our tests may 
% cover only a small portion of the \ac{SUT}. Therefore, the model may not reflect the full behaviour. 
% To mitigate this issue, we are working on an active approach similar to traditional active automata learning~\cite{DBLP:conf/sfm/SteffenHM11}.
% For that, we start with a small set of tests, learn a model, and then perform tests to find counterexamples to equivalence
% between \ac{SUT} and model. If we find counterexamples, we add these to our traces and refine the model via genetic programming. 
% This is iterated as long as we find counterexamples. Such an approach would improve performance as we could 
% start with a lower number of tests and by generating tests from approximate models we could increase coverage of the \ac{SUT}.

Assuming output urgency helps to approximate equivalence checks by ``testing'' candidate 
automata during learning. However, such models do not allow for uncertainty with respect to output timing. Relaxing this 
limitation represents an important next step. We are also currently working on this topic. 
% If we, e.g., change the conformance relation from equivalence to trace inclusion, we might simply 
% create a non-deterministic model with one location and self-loops for all actions without constraints on timing.
% The \ac{SUT}'s traces would be included. Such situations need to be avoided, e.g., by awarding negative fitness. 

We demonstrated that \tap{} can be genetically programmed, i.e. their structure is amenable to iterative refinement via mutation
and crossover. Therefore, we could apply the same approach, but base the fitness evaluation 
on model checking by adapting the technique presented by Katz and Peled~\cite{DBLP:journals/sttt/KatzP17}, to synthesise
%\tap{} satisfying some properties. For that we could, e.g., adapt the technique presented by Katz and Peled~\cite{DBLP:journals/sttt/KatzP17} 
\tap{} satisfying some properties. %For that we could, e.g., adapt the technique from~\cite{DBLP:journals/sttt/KatzP17}.
% to timed systems.  
This would enable learning a black-box system, which may contain errors, and synthesising a
controller ensuring that those errors do not lead to observable system failures.%  -- entirely via genetic programming. 

% \subsubsection{Active Learning.}

% \subsubsection{Expressiveness.}

\section*{Acknowledgment}
The work of B.\,Aichernig and M.\,Tappler has been carried out as part of the TU Graz LEAD project ``Dependable Internet of Things in Adverse Environments''.
The work of K.\,Larsen and F.\,Lorber has been conducted within the ENABLE-S3 project that has received funding from the ECSEL Joint Undertaking under grant agreement no.\,692455. 
This joint undertaking receives support from the European Union's Horizon 2020 research and innovation programme and Austria, Denmark, Germany, Finland, Czech Republic, Italy, Spain, Portugal, Poland, Ireland, Belgium, France, Netherlands, United Kingdom, Slovakia, Norway.
We would like to thank student Andrea Pferscher for her help in implementing the demonstrator.
\bibliographystyle{plain}
\bibliography{references}

\end{document}